\documentclass[floatfix,onecolumn,showpacs]{revtex4}

\usepackage{graphics,graphicx}
\usepackage{epsfig}
\usepackage{amsmath}
\usepackage{amstext}
\usepackage{amssymb}
\usepackage{dcolumn}

\newcommand{\beq}{\begin{eqnarray}}
\newcommand{\eeq}{\end{eqnarray}}
\begin{document}

\title{Reply to Comment on ``Towards a large deviation theory for strongly correlated systems"}

\author{Guiomar Ruiz$^{1,2}$  and Constantino Tsallis$^{1,3}$}
\affiliation{$^1$Centro Brasileiro de Pesquisas Fisicas and \\National Institute of Science and Technology for Complex Systems,
Rua Xavier Sigaud 150, 22290-180 Rio de Janeiro-RJ, Brazil\\
$^2$ Departamento de Matem\'{a}tica Aplicada y Estad\'{\i}stica, Universidad Polit\'{e}cnica de Madrid, Pza. Cardenal Cisneros s/n, 28040 Madrid, Spain\\
$^3$Santa Fe Institute, 1399 Hyde Park Road, Santa Fe, NM 87501, USA}

\begin{abstract}

The paper that is commented by Touchette contains a computational study which opens the door to a desirable generalization of the standard large deviation theory (applicable to a set of $N$ nearly independent random variables) to systems belonging to a special, though ubiquitous, class of strong correlations. It focuses on three inter-related aspects, namely (i) we exhibit strong numerical indications which suggest that the standard exponential probability law is asymptotically replaced by a power-law as its dominant term for large $N$; (ii) the subdominant term appears to be consistent with the $q$-exponential behavior typical of systems following $q$-statistics, thus reinforcing the thermodynamically extensive entropic nature of the exponent of the $q$-exponential, basically $N$ times the $q$-generalized rate function; (iii) the class of strong correlations that we have focused on corresponds to attractors in the sense of the Central Limit Theorem which are $Q$-Gaussian distributions (in principle $1 < Q < 3$), which relevantly differ from (symmetric) L\'evy distributions, with the unique exception of Cauchy-Lorentz distributions (which correspond to $Q = 2$), where they coincide, as well known. In his Comment, Touchette has agreeably discussed point (i), but, unfortunately, points (ii) and (iii) have, as we detail here, visibly escaped to his analysis. Consequently, his conclusion claiming the absence of special connection with $q$-exponentials is unjustified.
\end{abstract}

 \maketitle

Before addressing in detail the Comment by Touchette \cite{Touchette2012} on our paper \cite{RuizTsallis2012}, let us describe the physical scenario within which we have undertaken a possible generalization of the standard large deviation theory (LDT). A standard many-body Hamiltonian system in thermal equilibrium with a ther- mostat at temperature $T$ is described by the Boltzmann-Gibbs (BG) weight, proportional to $e^{-\beta {\cal H}_N} = e^{-\beta [{\cal H}_N/N]N}$, where ${\cal H}_N$ is the $N$-particle Hamiltonian, and $\beta \equiv 1/k_B T$. For standard Hamiltonian systems (typically involving short-range interactions and an ergodic behavior), the total energy is {\it extensive}. Consequently, the quantity $[{\cal H}_N/N]$ scales with $N$,  analogously to a (thermodynamically) {\it intensive} variable. This is to be compared with the LDT probability $P(N) \sim e^{- r_1N}$, where the rate function $r_1$ (the meaning of the subindex $1$ will soon become clear) is related to a BG entropic quantity {\it per particle}, and plays a role analogous to $\beta [\cal {H}_N/N]$ (we remind that, for such standard systems, $\beta$ is an intensive variable).

If now we focus on say a $d$-dimensional classical system involving two-
body interactions whose potential asymptotically decays at long distance $r$
like $-A/r^{\alpha}$ $(A > 0; \alpha \ge 0)$, the canonical BG partition function converges
whenever the potential is integrable, i.e. for $\alpha/d > 1$ (short-range interactions), and diverges whenever it is nonintegrable, i.e. for $0 \le \alpha/d \le 1$ (long-range interactions). The use of the BG weight becomes unjustified [``illusory" in Gibbs words \cite{Gibbs1902} for say Newtonian gravitation, which in the present notation corresponds to $(\alpha,d)=(1,3)$, hence $\alpha/d=1/3$]
in the later case because of the divergence of the BG partition function. We might therefore expect the emergence of some function
$f({\cal H}_N)$ different from the exponential one, in order to describe some specific stationary (or quasi-stationary) states differing from thermal equilibrium. The Hamiltonian ${\cal H}_N$ generically
scales like $N{\tilde N}$ with ${\tilde N} \equiv \frac{ N^{1-\alpha/d}-1}{1- \alpha/d} \equiv \ln_{\alpha/d} N$ (with the {\it $q$-logarithmic function} defined as $\ln_q z \equiv \frac{z^{1-q}-1}{1-q}; \,z > 0; \, \ln_1 z = \ln z$). Notice that ($N \to \infty$)
${\tilde N} \sim N^{1-\alpha/d}/(1-\alpha/d)$ for $1 \le \alpha/d < 1$,
${\tilde N} \sim \ln N$ for $\alpha/d = 1$,
and ${\tilde N} \sim1/(\alpha/d -1)$ for $\alpha/d > 1$. The particular case $\alpha = 0$ yields ${\tilde N} \sim  N$, thus recovering the usual prefactor of Mean Field theories. The quantity $\beta {\cal H}_N$ can be rewritten as $[(\beta {\tilde N}){\cal H_N} /(N{\tilde N})]N = [{\tilde \beta}{\cal H}_N /(N{\tilde N})]N$, where ${\tilde \beta} \equiv \beta {\tilde N} \equiv 1/k_B{\tilde T} = {\tilde N}/k_B T$ plays the role of an intensive variable. The correctness of all these scalings has been profusely verified in various kinds of thermal \cite{thermal}, diffusive \cite{diffusion} and geometrical (percolation) \cite{geometrical} systems (see also \cite{Tsallis1988,Tsallis2009}). We see that, not only for the usual case of short-range interactions but also for long-range ones, $[{\tilde \beta}{\cal H}_N/(N{\tilde N})]$ plays a role analogous to an intensive variable.
The {\it $q$-exponential function} $e_q^z \equiv [1 + (1 - q) z]^{\frac{1}{1-q}}$ ($e_1^z = e^z$) (and its associated $q$-Gaussian \cite{reminder}) has already emerged, in a considerable amount of nonextensive and similar systems (see \cite{TamaritCannasTsallis1998,AnteneodoTsallis1997,TirnakliTsallisLyra1999,TsallisAnjosBorges2003,GellMannTsallis2004,RodriguezSchwammleTsallis2008,AndradeSilvaMoreiraNobreCurado2012,TirnakliJensenTsallis2011,PlastinoPlastino1995,pedron,AnteneodoTsallis2003,PluchinoRapisardaTsallis2007,AnteneodoTsallis1998,UpadhyayaRieuGlazierSawada2001,ThurnerWickHanelSedivyHuber2003,DanielsBeckBodenschatz2004,BurlagaVinas2005,DouglasBergaminiRenzoni2006,LiuGoree2008,CMS,TirnakliBeckTsallis2007,NobreRegoMonteiroTsallis2012}  among others), as the appropriate generalization of the exponential one (and its associated Gaussian). Therefore, it appears as rather natural to conjecture that, in some sense that remains to be precisely defined, the LDT expression $e^{-r_1N}$ becomes generalized into something close to $e_q^{-r_q N}$ ($q \in {\cal R}$), where the generalized function rate $r_q$ should be some generalized entropic quantity {\it per particle}.
Let us stress a crucial point: we are {\it not} proposing for long-range interactions, and other nonstandard systems, something like $e_q^{-r_q N^\gamma}$ with $\gamma \ne 1$, but we are expecting instead $\gamma =1$, i.e., the {\it extensivity} of the total $q$-generalized entropic form to still hold \cite{extensive}, in order to be consistent with many other related results (e.g., \cite{Tsallis2009,UmarovTsallisSteinberg2008,VignatPlastino2007,Hilhorst2010}). We shall soon see that this important assumption indeed appears to be verified in the model, characterized by $(Q,\gamma,\delta)$ , that we numerically studied in \cite{RuizTsallis2012}.

\begin{figure}
\begin{center}
\includegraphics[width=14.0cm]{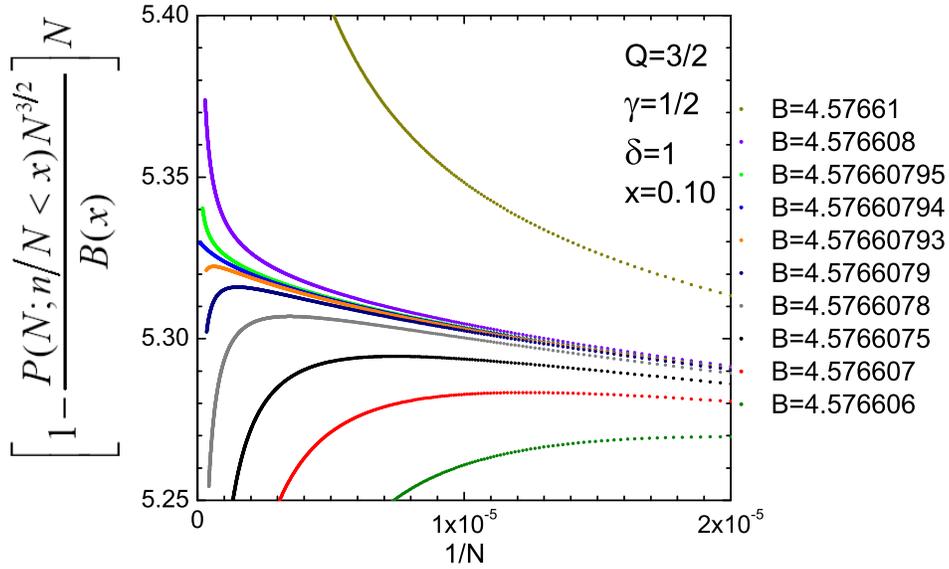}
\end{center}
\vspace{-1cm}
\caption{Detailed numerical verification of the conjecture given by Eq. (\ref{generalization}) in {\it one} of the 10 ``lines" (the bottom one, to be more precise) observed in Fig. \ref{generalization} for $x=0.1$ . This procedure enables a high precision numerical determination of $B(x)$ for any chosen value of $x$. For a given $(Q,\gamma,\delta)$ model, the value of $B(x)$ is one and the same for all the ``lines" associated with a given value of $x$. Not so for $C(x)$: indeed, for fixed $x$, we observe the existence of a set of values for $C(x)$ which we note $\{C_j(x)\}$, with $j=1,2,\dots ,j_{max}$ (in the present illustration $j_{max}=10$). The finiteness of the set $\{C_j(x)\}$ here and in Fig. \ref{method2} means that the corrections to the $N^{-\eta}$ power in Eq. (\ref{asymptotic}) are of the $1/N$ order. The finiteness that we observe (here and in Fig. \ref{method2}) in the slopes at the origin means that the next corrections are of the $1/N^2$ order. In this example, we have run $N$ up to $11 \times 10^6$.
}
\label{method}
\end{figure}

\begin{figure}
\begin{center}
\includegraphics[width=11cm]{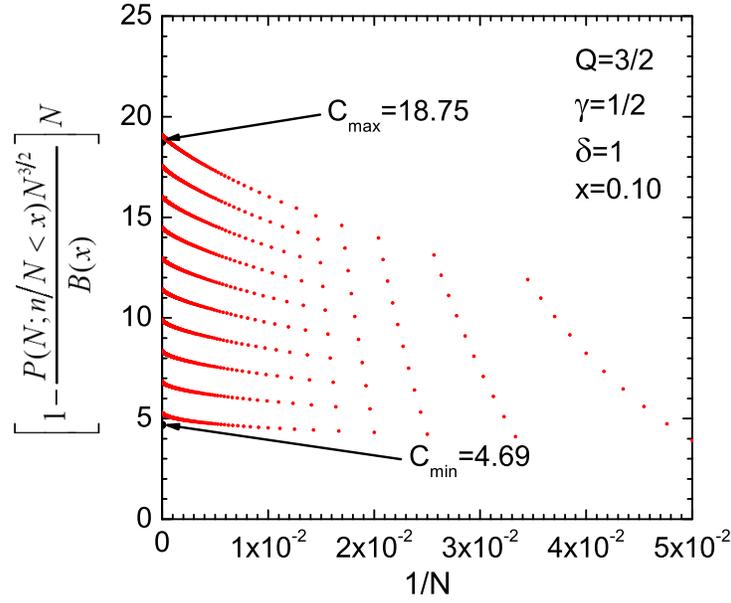}
\end{center}
\vspace{-1cm}
\caption{Numerical verification of the conjecture given by Eq. (\ref{generalization}), $N$ running up to $11 \times 10^6$. For fixed $x$, $B(x)$ is unique, whereas $C(x)$ corresponds to a set of values $\{C_j\}$ ($j=1,2, \dots , j_{max}$), where $j_{max}$ depends on $x$. We also see that the next correction is of the type $1/N^2$. The upper and lower values $C_{min}$ and $C_{max}$ that are indicated by the arrows precisely correspond to the upper and lower bounds $r_q^{(lower\,bound)}(x)$ and  $r_q^{(upper\,bound)}(x)$ such that all present numerical results are
within two $q$-exponentials, as indicated in Figs. \ref{comparison} and \ref{comparisonqlog}.
}
\label{method2}
\end{figure}

Let us start by exhibiting that its $N\to\infty$ LDT asymptotic behavior numerically satisfies
\begin{equation}
P(N; n/N <x) \sim \frac{B(x)}{N^\eta} \Bigl[1-\frac{C(x)}{N} \Bigr]  \;\;\;(B(x) > 0; \, C(x) >0)\,,
\label{asymptotic}
\end{equation}
with $ \eta \equiv \frac{1}{q-1}= \frac{\gamma (3-Q)}{Q-1}> 0$ \cite{RuizTsallis2012}. This implies the existence of a generically positive {\it finite} $B(x)$ such that
\begin{equation}
\lim_{N \to\infty}\Bigl[1 - \frac{P(N; n/N < x)N^\eta}{B(x)}\Bigr]N=C(x),
\label{generalization}
\end{equation}
$C(x)$ being a generically positive {\it finite} number
for all values of $x$ different from 1/2. This is indeed verified, as exhibited in Fig. \ref{method} and \ref{method2}. More precisely, we verify for fixed $(Q,\gamma,\delta)$ that $B(x)$ is {\it unique} for any given $x$, whereas $C(x)$ is in fact a set of values, noted $\{C_j(x)\}$, with $j=1,2, \dots, j_{max}$ (the value of $j_{max}$ depends on $x$; for example, we can see that, for the illustration exhibited in Fig. \ref{method2}, $j_{max}=10$ for $x=0.1$).
Let us emphasize that the $1/N$ correction to the power law $1/N^\eta$ in (\ref{asymptotic}) is consistent with the total entropy of the system always being {\it extensive} in the thermodynamical sense.

Let us next check the conjecture made in \cite{RuizTsallis2012}, namely that $P(N; n/N < x)$ is, for $q>1$, well approached by
\begin{eqnarray}
P(N; n/N <x) &=& a(x) e_q^{-r_q(x) N} \nonumber \\
&=&\frac{a(x)}{[1+(q-1)r_q(x)]^{\frac{1}{q-1}}} e_q^{-\frac{r_q(x)}{1+(q-1)r_q(x)}(N-1)}       \nonumber \\
&=&e_q^{-\Bigl\{\frac{r_q(x)}{[a(x)]^{q-1}}N+ \frac{1-[a(x)]^{q-1}}{(q-1)[a(x)]^{q-1}}\Bigr\}} \nonumber \\
&=& \frac{a(x)}{[1+ (q-1) r_q(x) N]^{\frac{1}{q-1}}} \nonumber \\
&=&\frac{a(x)}{[(q-1)r_q(x) N]^{\frac{1}{q-1}}} \nonumber \\
&\times& \Bigl[1-\frac{1}{(q-1)^2 r_q(x) N}+ \frac{q}{2(q-1)^4 (r_q(x) N)^2} - \frac{q(2q-1)}{6(q-1)^6 (r_q(x) N)^3}  + \dots        \Bigr] \nonumber \\
&=&\frac{a(x)}{[(q-1)r_q(x) N]^{\frac{1}{q-1}}} \nonumber \\
&\times& \Bigl[1-\frac{1}{(q-1)^2 r_q(x) N}+ \sum_{m=2}^{\infty}(-1)^m \frac{ q(2q-1)\dots [(m-1)q-(m-2)]}{m!(q-1)^{2m}(r_q(x)N)^m} \Bigr] \,.
\label{eq1}
\end{eqnarray}
By identifying this expansion with Eq. (\ref{asymptotic}) we obtain
\begin{equation}
a_j(x)=B(x)[(q-1)r_q^{(j)}(x)]^{\frac{1}{q-1}} \;\;\;(j=1,2, \dots , j_{max}) \,,
\label{identity1}
\end{equation}
and
\begin{equation}
r_q^{(j)}(x)=\frac{1}{(q-1)^2C_j(x)}     \;\;\;(j=1,2, \dots , j_{max})   \,.
\label{identity2}
\end{equation}

\begin{figure}
\begin{center}
\includegraphics[width=12cm]{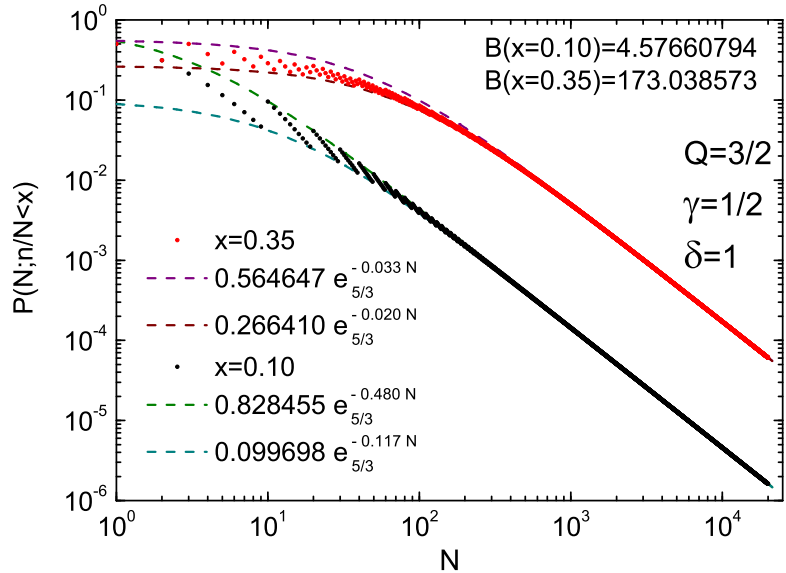}
\end{center}
\vspace{-1cm}
\caption{Comparison of our numerical data (dots) with $a(x)e_q^{-r_q N}$, where $(a(x),r_q(x))$ have been calculated from $(B(x),C(x))$ by using Eqs. (\ref{identity1}) and (\ref{identity2}). The values for $C(x)$ that have been used are those indicated by arrows in Fig. \ref{method2}. Two values for $x$, namely $x=0.10$ and $x=0.35$, have been illustrated here.
}
\label{comparison}
\end{figure}

\begin{figure}
\begin{center}
\includegraphics[width=13cm]{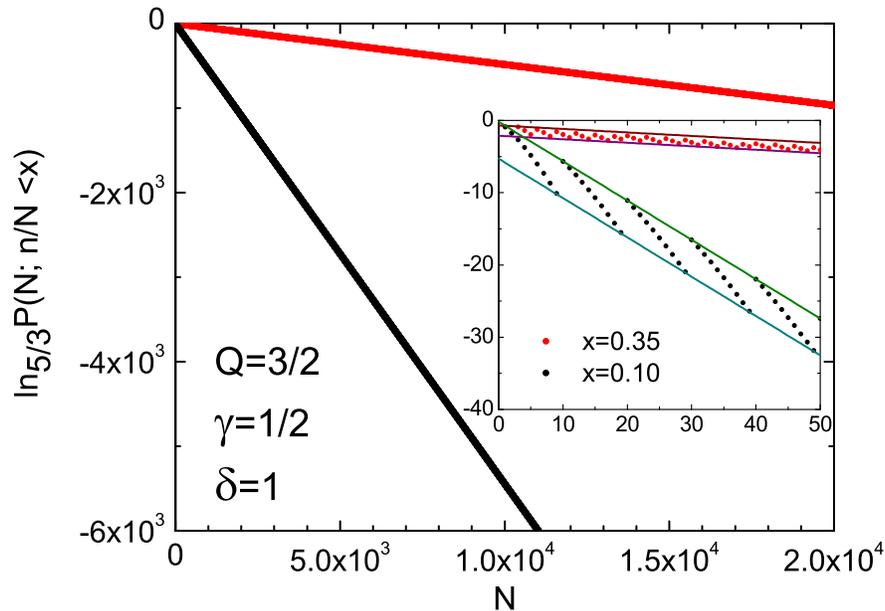}
\end{center}
\vspace{-1cm}
\caption{The same data of Fig. \ref{comparison} in ($q$-log)-linear representation. Let us stress that the {\it unique} asymptotically-power-law function which provides straight lines {\it at all scales} of a ($q$-log)-linear representation is the $q$-exponential function. The inset shows the results corresponding to $N$ up to 50.
}
\label{comparisonqlog}
\end{figure}

Since $B(x)$ and $\{C_j(x)\}$ are numerically known, we can easily calculate $\{a_j(x)\}$ and $\{r_q^{(j)}(x)\}$ by using Eqs. (\ref{identity1}) and (\ref{identity2}).
Knowing these, we calculate $a_j(x)\,e_q^{-r_q^{(j)}(x)N}$ ($j=1,2, \dots , j_{max}$) and compare with our numerical data.
We then bound our numerical results from both below and above (see Figs. \ref{comparison} and \ref{comparisonqlog} for illustrations). More precisely, for each value of $x$, we have adopted two values, noted $C_{lower\,bound}(x)$ and  $C_{upper\,bound}(x)$, such that $q$-exponential upper and lower bounds for the entire set of numerical values for $P(N; n/N <x)$ are obtained.  These $ C_{lower\,bound}(x)$ and $ C_{upper\,bound}(x)$ values turn out to be comparable to the corresponding set $\{C_j(x) \}$ (see Fig. \ref{method2}).
In other words, we obtain the values of $C_{lower\,bound}(x)$, $C_{upper\,bound}(x)$, $r_q^{(upper\,bound)}(x)$ and $r_q^{(lower\,bound)}(x)$, consistent with  Eq. (\ref{identity2}).
By introducing these values in Eq. (\ref{identity1}) we obtain $a_q^{(upper\,bound)}(x)$ and $a_q^{(lower\,bound)}(x)$, and verify that, in the model studied in \cite{RuizTsallis2012}, $\forall x$, $\forall N$,
\begin{eqnarray}
&&B(x)[(q-1)r_q^{(lower\,bound)}(x)]^{\frac{1}{q-1}}  e_q^{-r_q^{(lower\,bound)}(x) \,N}  \nonumber \\
&=&\frac{B(x)}{N^{\frac{1}{q-1}}} \Bigl[1-\frac{1}{(q-1)^2 \, r_q^{(lower\,bound)}(x)}\frac{1}{N} + o(1/N^2)    \Bigr]   \nonumber \\
&&\le P(N; n/N <x)  \le  \nonumber \\
&&B(x)[(q-1)r_q^{(upper\,bound)}(x)]^{\frac{1}{q-1}} e_q^{-r_q^{(upper\,bound)}(x) \,N} \nonumber \\
&=&\frac{B(x)}{N^{\frac{1}{q-1}}} \Bigl[1-\frac{1}{(q-1)^2 \, r_q^{(upper\,bound)}(x)}\frac{1}{N} + o(1/N^2)    \Bigr]   \,.
\label{sandwich}
\end{eqnarray}



\begin{figure}
\begin{center}
\includegraphics[width=12cm]{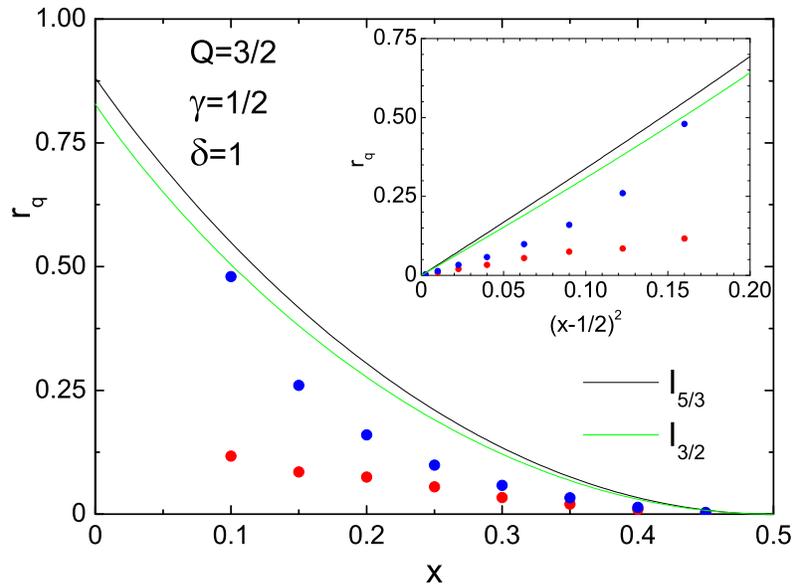}
\end{center}
\vspace{-1cm}
\caption{$x$-dependences of the lower and upper bounds for the rate function $r_q(x)$ obtained from $q$-exponential fittings of the numerical data (see Figs. \ref{comparison} and \ref{comparisonqlog}). The analytical curves $I_{3/2}(x)$ and $I_{5/3}(x)$ are included for comparison. The inset exhibits the quasi-parabolic behavior at both sides of $x=1/2$.
 }
\label{fig3}
\end{figure}

We may summarize the above considerations by conjecturing that, for all strongly correlated systems which have $Q$-Gaussians ($Q>1$) as attractors in the sense of the central limit theorem (see \cite{UmarovTsallisSteinberg2008}), a model-dependent set $[q>1, B(x)>0, r_q^{(lower\,bound)}(x)>0, r_q^{(upper\,bound)}(x)>0]$ might exist such that $P(N; n/N <x)$ generically satisfies inequalities  (\ref{sandwich}). In our present example, this set depends on $(Q,\gamma,\delta)$. Typical values of $[r_q^{(lower\,bound)}(x), r_q^{(upper\,bound)}(x)]$ are illustrated in Fig. \ref{fig3} and compared with $q$-generalized entropic quantities.


Touchette mentions Kaniadakis' $\kappa$-logarithm and $\kappa$-exponential \cite{Kaniadakis2001} as an alternative to the $q$-exponential and $q$-logarithm herein conjectured. Let us address this point through the definition
\begin{equation}
\ln_\kappa z\equiv \frac{z^\kappa -z^{-\kappa}}{2\kappa} = \frac{1}{2}\Bigl(\ln_q z - \ln_q \frac{1}{z} \Bigr)                 \;\;\;(q=1+\kappa)
\label{eq3}
\end{equation}
(Notice a misprint in the definition of the $\kappa$-logarithm appearing in Touchette's Comment). It straightforwardly follows the asymptotic series
\begin{eqnarray}
e_\kappa^{-r_\kappa N} &=&
\frac{1}{
[
\sqrt{1+(\kappa r_k N)^2}+\kappa r_k N
]^{1/\kappa} } \nonumber \\
 &=&\frac{1}{[2\kappa \,r_\kappa N]^{1/\kappa}}
 \Bigl[1-\frac{1}{4
\kappa^3(r_{\kappa} N )^2 } \nonumber \\
 &&\qquad +\sum_{m=2}^{\infty}(-1)^{m}\frac{[(m+1)\kappa+1]\cdot[(m+2)\kappa+1]
\cdots[(2m-1)\kappa+1]}{2^{2m}m!\kappa^{3m}(r_{\kappa} N )^{2m}}     \Bigr]\nonumber\\
\label{eq4}
\end{eqnarray}
The dominant term is a power-law, and at this approximation it is trivially as admissible as virtually any other power-law. However, we verify a highly  meaningful discrepancy with the $q$-exponential function, namely that its subdominant correction is in $1/N^2$, instead of $1/N$. This fact excludes the $\kappa$-exponential function as an adequate one for the present purpose; indeed, it cannot satisfactorily reproduce the results exhibited in Figs. \ref{comparison} and \ref{comparisonqlog} .

A point remains to be discussed. The L\'evy-Gnedenko theorem concerns sums of infinitely many {\it independent} (or nearly independent, in a specific sense) random variables, whereas the 2008 $Q$-central limit theorem \cite{UmarovTsallisSteinberg2008}  concerns sums of infinitely many {\it strongly correlated} variables within a specific class. The first case corresponds to divergent standard variance, whereas the second one concerns finite ${\bar Q}$-variance (${\bar Q} = 2Q - 1$; see details in \cite{UmarovTsallisSteinberg2008}). The attractors for the former case are L\'evy distributions, whereas those for the latter are $Q$-Gaussians. Both classes have long tails. For the L\'evy distributions, the decay is slower than $1/|x|^3$ and faster than $1/|x|$; for the $q$-Gaussians, any power-law faster than $1/|x|$ is admissible. It is known that they have this and other relevant differences. They always differ excepting for an unique case, which happens to be precisely the case focused on by Touchette, i.e. $Q=2$, namely the Cauchy-Lorentz distribution (named after Cauchy by mathematicians, and after Lorentz by physicists). They can be simply thought as having $r_1(x) = 0$, which, as acknowledged by Touchette, is not particularly enlightening. But they can be also thought in a much more interesting way, namely as having $r_2(x)$ different from zero, which neatly illustrates the usefulness of the approach adopted in \cite{RuizTsallis2012}. In fact, it is well known that $Q = 2$ is a highly peculiar case within the interval $1 < Q < 3$. For example, the anomalous diffusion coefficient in the nonlinear Fokker-Planck equation known as the Porous Medium Equation and discussed in \cite{PlastinoPlastino1995} changes its sign precisely at $Q=2$ (see also \cite{pedron}).
The fact that, for $Q = 2$, $r_1 = 0$ whereas $r_q \ne 0$ is totally analogous to a variety of dissipative one-dimensional maps whose Lyapunov exponent vanishes at the edge of chaos. In such cases, the use of the nonadditive entropy $S_q$ instead of the BG one makes the discussion much richer since it enables a simple quantitative characterization of the nonlinear dynamical behavior (by generalizing the standard exponential sensitivity to the initial conditions when the maximal Lyapunov exponent is strictly positive to the $q$-exponential form at the edge of chaos, when the maximal Lyapunov exponent vanishes). This has been verified both analytically and numerically in very many cases \cite{TirnakliTsallisLyra1999}.

Let us conclude by saying that point (i) of the present Abstract is agreeably discussed in Touchette's Comment, but a neat analysis of the important points (ii) and (iii) is notoriously absent in his paper. In other words, the $q$-exponential ansatz proposed in \cite{RuizTsallis2012} for (asymptotically) generalizing the standard LDT remains (either exactly or approximatively: see the quantity (\ref{generalization}), expected to be finite, and the inequalities (\ref{sandwich})) as a very strong candidate for a wide class of systems whose elements are strongly correlated. This fact may be seen as a strong indication that, consistently with other results available in the literature (see \cite{thermal,diffusion,geometrical,Tsallis2009,GellMannTsallis2004,extensive}), the total entropy remains {\it extensive} (i.e., thermodynamically admissible) even in nonstandard cases where the BG entropy fails to be extensive. Any analytical results along these or similar lines would obviously be highly interesting and welcome.

\section*{Acknowledgments}
We acknowledge
partial financial support by DGU-MEC (Spanish Ministry of Education) through Project PHB2007-0095-PC, and by CNPq, Faperj and Capes (Brazilian agencies). \\


\end{document}